\documentclass[sigconf,nonacm]{acmart}

\usepackage{makecell}
\usepackage{graphicx}
\usepackage{xspace}
\usepackage{enumitem}
\usepackage{bbding} 
\AtBeginDocument{%
  }

\renewcommand\footnotetextcopyrightpermission[1]{}

\newcommand{\tool}{{\sc \texttt{ItoV}}\xspace}

\acmSubmissionID{975}

\begin{document}
\sloppy
\title{ItoV: Efficiently Adapting Deep Learning-based Image Watermarking to Video Watermarking}

\author{
 Guanhui Ye, ~~Jiashi Gao, ~~Yuchen Wang, ~~Liyan Song,~~Xuetao Wei 
\\
	{Southern University of Science and Technology}\\
}

\renewcommand{\shortauthors}{Ye et al.}

\begin{abstract}
Robust watermarking tries to conceal information within a cover image/video imperceptibly that is resistant to various distortions. Recently, deep learning-based approaches for image watermarking have made significant advancements in robustness and invisibility. However, few studies focused on video watermarking using deep neural networks due to the high complexity and computational costs. Our paper aims to answer this research question: Can well-designed deep learning-based image watermarking be efficiently adapted to video watermarking?
 Our answer is positive. First, we revisit the workflow of deep learning-based watermarking methods that leads to a critical insight: temporal information in the video may be essential for general computer vision tasks but not for specific video watermarking. Inspired by this insight, we propose a method named \tool for efficiently adapting deep learning-based \textbf{I}mage watermarking \textbf{to} \textbf{V}ideo watermarking. Specifically, \tool merges the temporal dimension of the video with the channel dimension to enable deep neural networks to treat videos as images. We further explore the effects of different convolutional blocks in video watermarking. We find that spatial convolution is the primary influential component in video watermarking and depthwise convolutions significantly reduce computational cost with negligible impact on performance. In addition, we propose a new frame loss to constrain that the watermark intensity in each video clip frame is consistent, significantly improving the invisibility. Extensive experiments show the superior performance of the adapted video watermarking method compared with the state-of-the-art methods on Kinetics-600 and Inter4K datasets, which demonstrate the efficacy of our method \tool.

\end{abstract}





\maketitle
\pagestyle{plain}

\section{Introduction}
With the rapid development of Internet video sharing services, video has gradually become the dominant multimedia content among Internet users. Many videos are published daily on platforms like YouTube and TikTok. Meantime, digital watermarking is widely used to safeguard multimedia content against copyright infringement. This method conceals unique watermark information within digital content, including images, videos, and audio files\cite{alenizi2017robust}.
There are three critical metrics to measure digital watermarking performance: capacity, invisibility, and robustness against various distortions. These metrics are in tension with each other, meaning that improving the performance of one metric will result in a decline in the performance of other metrics. Our primary goal in this paper is to develop a digital video watermarking method that achieves higher robustness and invisibility. 


The earliest research on digital watermarking, known as Least Significant Bits (LSB) \cite{van1994digital}, involved encoding a secret message onto the least significant bits of image pixels. To improve the performance, traditional methods turned to focus on the transfer domain, such as the DFT domain \cite{ruanaidh1996phase}, DCT domain \cite{hamidi2018hybrid}, and DWT domain \cite{guo2003digital}. These methods adjusted the watermark embedding strategy based on the data distribution. However, traditional watermarking methods rely heavily on shallow hand-craft features, which require careful design and cannot fully utilize the redundant information of cover videos. As a result, traditional methods struggle to achieve robustness against various distortions simultaneously.

\begin{figure}[ht]
  \centering
  \includegraphics[width=\linewidth]{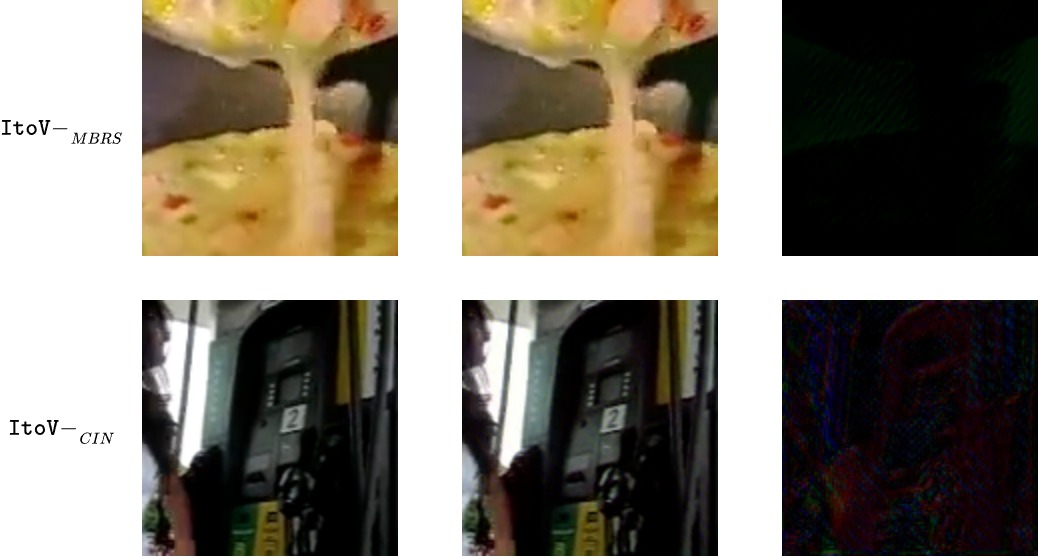}
  \caption{We utilize our proposed method \tool to adapt two state-of-the-art image watermarking methods, MBRS \cite{jia2021mbrs} and CIN \cite{ma2022towards}, to the video watermarking. Here are some visual samples of video frames from test datasets. From left to right are: the cover video frames $V_c$, the watermarked video frames $V_w$, and the residual signals between them R = $|V_w - V_c|$. The residual signals are five times magnified for visualization.
  }
    \label{fig:visual}
\end{figure}

\begin{figure*}[ht]
  \centering
  \includegraphics[width=0.91\linewidth]{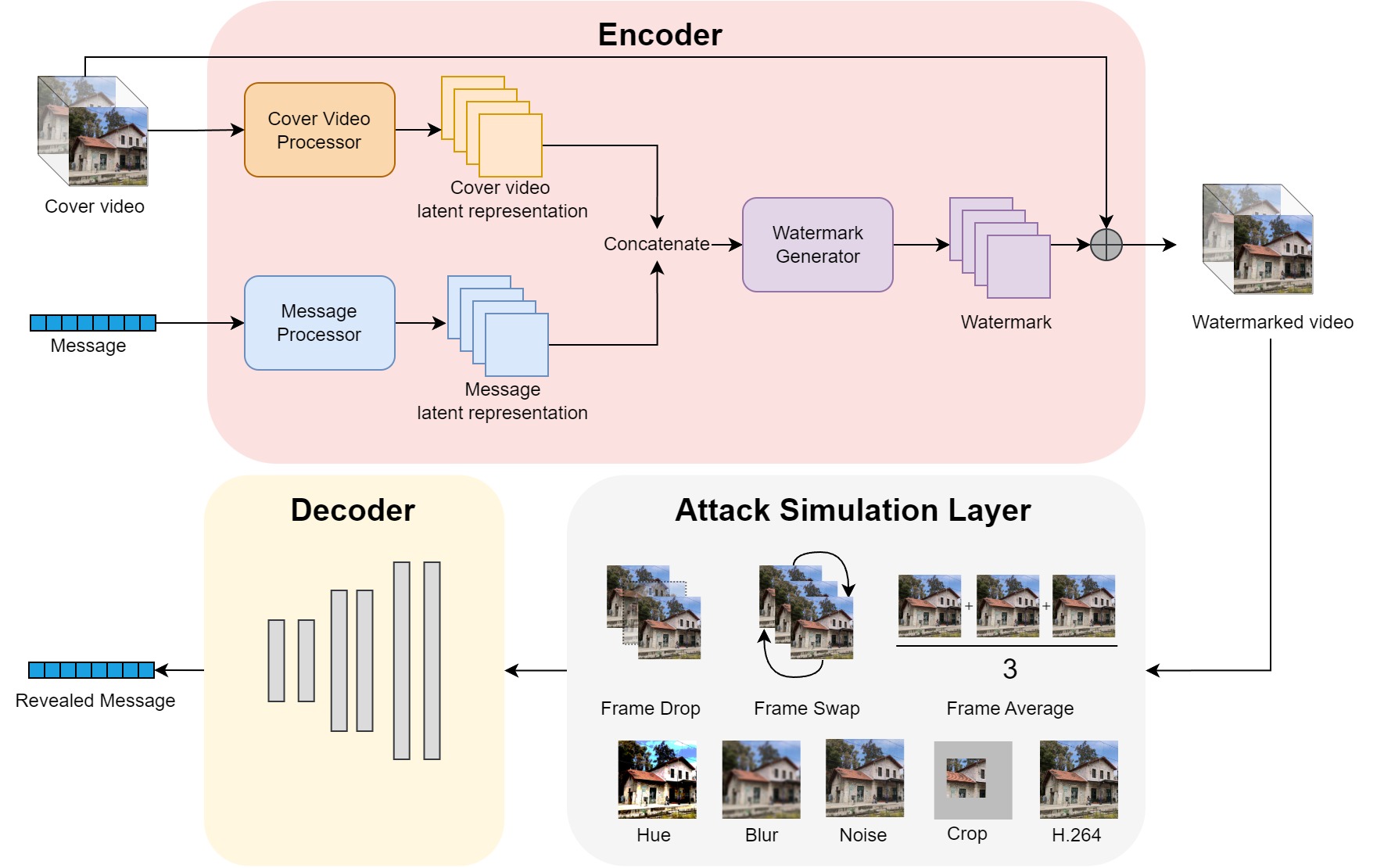}
  \caption{The \tool framework, which consists of three main parts: the encoder, the attack simulation layer, and the decoder}
    \label{fig:framework}
\end{figure*}

Deep learning-based image watermarking has recently achieved remarkable success \cite{fang2022pimog,ma2022towards,jia2021mbrs}, performing significantly better than traditional watermarking methods. However, there is still limited research on deep learning-based video watermarking techniques due to the complexity of neural networks for videos and the high computational costs required for training. Although we can treat video frames as images and directly use digital image watermarking methods to embed watermarks in them, such a straightforward approach does not allow the neural network to learn useful features based on the characteristics of video data, making it challenging to resist various distortions targeted at videos. We can also use 3D convolution instead of 2D convolution in image watermark networks to achieve video watermarking. However, this change results in a significant increase in computational cost.

To address these limitations, we revisit the process of embedding and extracting watermarks in images/videos via neural networks. We find that embedding a watermark in either images or videos is a similar task for neural networks, as they embed watermarks into images or videos based on pixel data distribution. Our key insight is that images and videos are just data with different shapes for watermarking neural networks, and the temporal information in videos is not essential for digital watermarking. To verify the effectiveness of our insight, we propose a method \tool to efficiently adapt deep learning-based \textbf{I}mage watermarking methods \textbf{to}  \textbf{V}ideo watermarking. Specifically, our \tool treats videos as images by merging the temporal and channel dimensions. Unlike adding watermarks independently to each video frame, we feed the video clip as a whole into the neural network, the results of which achieve remarkable robustness against various video distortions such as H.264 compression, frame average, and screen recording.

Furthermore, we incorporate two spatiotemporal convolutional blocks, 3D \cite{ji20123d} and (2+1)D \cite{tran2018closer}, in our network to explore how temporal convolution affects digital watermarking tasks in videos. Our findings indicate that spatial convolution is the primary influential component in video watermarking. Additionally, we observe that spatiotemporal convolutional networks can easily lead to over-parameterization, which increase expensive computational cost and the risk of overfitting. Therefore, we apply depthwise convolutions \cite{tran2019video,howard2017mobilenets} to the video watermarking networks to reduce the computational complexity of the method.

In summary, the main contributions of this paper are:

\begin{itemize}[leftmargin=0.7cm, itemindent=0cm]

\item  We revisit the process of deep learning-based watermarking methods and obtain one key insight that temporal information in the video may be essential for general computer vision tasks but not for specific video watermarking. Namely, a network that only comprehends spatial information in videos can effectively fulfill the task of video watermarking. 

\item  Inspired by the critical insight above, we propose a method called \tool that merges the temporal dimension of videos into the channel dimension, which efficiently adapts deep learning-based image watermarking methods to video watermarking. In addition, we design a frame loss to control consistent watermark strength across all video frames, which significantly improves invisibility.

\item We conduct a study on the impact of various convolutional blocks in video watermarking. Our results suggest that spatial convolution is the primary influential component instead of temporal convolution, while depthwise convolutions significantly decrease computational cost with negligible impact on performance.

\item We conduct extensive experiments to demonstrate that our method \tool achieves higher robustness and invisibility than state-of-the-art video watermarking methods under various distortions.

\end{itemize}

\section{Related Work}

\subsection{Deep learning-based watermarking}
Digital watermarking is a popular technique for copyright protection of content. 
In recent years, deep learning-based methods have demonstrated great invisibility and robustness against various distortions simultaneously \cite{zhu2018hidden,jia2021mbrs,ma2022towards,liu2019novel,ahmadi2020redmark}, which is due to the powerful feature extraction ability of deep neural networks. Zhu et al. \cite{zhu2018hidden} was the first to propose an end-to-end deep learning-based method for watermarking. The primary architecture follows an encoder-noise layer-decoder structure, like an auto-encoder. Additionally, a discriminator was employed to reduce the visibility of watermarks for human perception. Furthermore, Ahmadi et al. \cite{ahmadi2020redmark} incorporated the domain transform technique with a neural network and utilized a strength factor to adjust the intensity of the watermark in the image. Some other work aimed to expand the application scope by focusing on complex and realistic distortions \cite{tancik2020stegastamp, fang2022pimog, wengrowski2019light}.

In order to achieve high robustness against expected distortions, the most common and effective approach is introducing an attack simulation layer between the encoder and decoder during the training. However, this method is unsuitable for non-differentiable distortions such as JPEG compression and H.264 compression. Hidden \cite{zhu2018hidden} introduced differentiable JPEG-Mask and JPEG-Drop to approximate real JPEG compression to address the limitation. However, the method still lacks robustness against real JPEG compression. Liu et al. \cite{liu2019novel} developed a two-stage separable framework to solve non-differentiable distortion problems.
Jia et al. \cite{jia2021mbrs} focused on improving robustness against JPEG compression by proposing a novel mini-batch of simulated and real JPEG compression training methods. Subsequently, Zhang et al.  \cite{zhang2021towards} proposed to adopt a forward attack simulation layer as a simple and effective method to improve the robustness of deep learning-based watermarking against non-differentiable distortions. 

Deep learning-based watermarking has been applied to videos \cite{luo2021dvmark, weng2019high, zhang2019robust} as well. Weng et al. \cite{weng2019high} introduced U-net \cite{chen2017deeplab} for video steganography and achieved excellent performance. However, they didn't consider video distortions. RivaGAN \cite{zhang2019robust} maintained both high invisibility and robustness against three common video distortions, but it ignored the non-differentiable H.264 compression, the most crucial distortion in the video. Luo et al. \cite{luo2021dvmark} developed a neural network that simulates H.264 compression to improve its robustness against real H.264 compression in DvMark. Previous work has demonstrated the potential of deep learning for digital watermarking.

\subsection{Convolutional neural networks for videos}
Although 3D convolutional neural networks(CNNs) have demonstrated superior performance over traditional methods in video watermarking, existing 3D networks typically require a significant amount of computing resources \cite{luo2021dvmark}. In addition, the training of 3D CNNs is more unstable and has a slow convergence speed \cite{carreira2017quo,liu20173d,tajbakhsh2016convolutional}. Hence, researchers are seeking efficient alternatives to 3D CNNs. Simonyan and Zisserman \cite{simonyan2014two} proposed the two-stream framework, which uses two 2D CNNs to learn the spatial and temporal information of videos, respectively. Tran et al. \cite{tran2018closer} improved the performance and efficiency of 3D convolutional layers by decomposing them into factorized convolutional layers. MobileNet \cite{howard2017mobilenets} introduced depthwise convolution to optimize the model size and computational cost for mobile applications. Tran et al. \cite{tran2019video} utilized depthwise convolutions to 3D networks to develop resource-efficient models. These convolutional blocks have been shown to generalize effectively for various tasks, such as action detection, temporal action localization, and hand gesture detection. However, no research has investigated the influence of different convolutional blocks on video watermarking.

\begin{figure*}[ht]
  \centering
  \includegraphics[width=\linewidth]{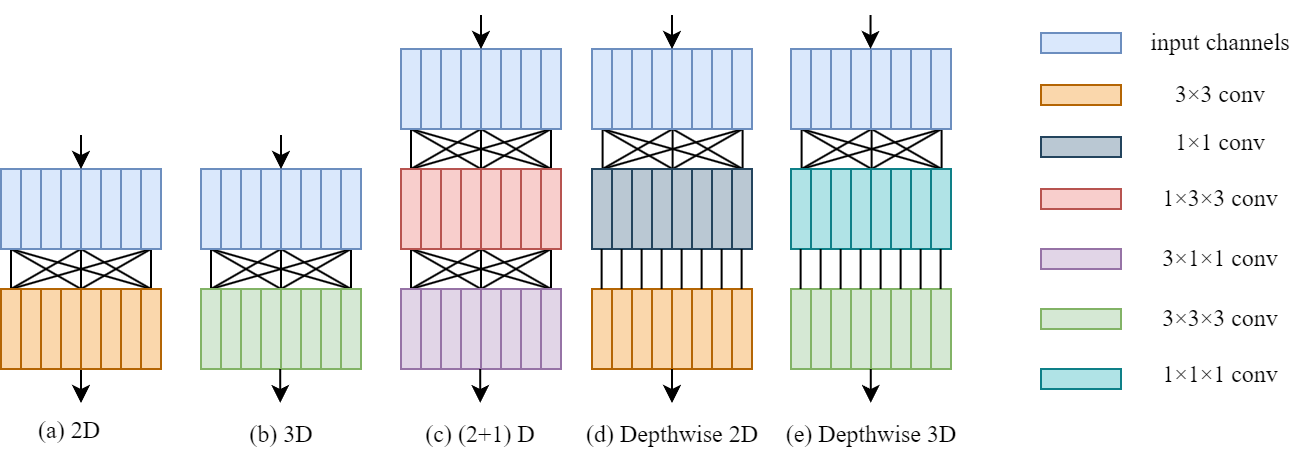}
  \caption{Convolutional blocks for video watermarking are considered in this work. (a) 2D convolution; (b) 3D convolution; (c) (2+1)D convolution; (d) Depthwise 2D convolution; (3) Depthwise 3D convolution.}
    \label{fig:conv}
\end{figure*}

\section{Proposed Method}


In this section, we discuss whether the temporal information of video is important for digital watermarking tasks. How to extract temporal information from videos has always been a focus of research in the field of video processing, as temporal information plays an essential role in making judgments on video understanding tasks.
For instance, in tasks related to action recognition, a video depicts the act of opening a box using two hands. Based on spatial information alone, we can infer that the video includes two hands and one box. However, in the absence of temporal information, it becomes challenging for the neural network to distinguish whether this is an instance of opening or closing a box. But for the embedding process of watermark tasks, the objective of neural networks is to understand the pixel distribution (spatial information) of the cover video so that messages can be embedded into the cover video without altering it as much as possible. Therefore, videos and images can be treated equally in digital watermarking since videos are just data with a different shape compared with images. Based on this insight, we propose a method \tool to efficiently adapt deep learning-based image watermarking techniques to the video. The neural network can treat the L frames in a video as channels and disregard their temporal information. This allows us to view these videos as reshaping the input 4D data tensors into 3D data tensors with dimensions of $\texttt{3}L\times H \times W$, in which \texttt{3} refers to the RGB channels, L is frame number, H is height, and W is width. Hence, current image watermarking methods can be directly applied to video watermarking.

The primary goal of deep learning-based video watermarking is to develop an end-to-end trainable architecture that demonstrates a notable level of robustness against various distortions. The framework is shown in Figure \ref{fig:framework}. It consists of three major components: (1) An encoder E with parameters $\theta_E$, which receives the cover video $V_c$ and the binary secret message $M$ $\in \{0,1\}^m$ of length m, and outputs the watermarked video $V_w$; (2) an attack simulation layer (ASL) receives the watermarked data $V_w$ and applies a random attack to generate the attacked video $V_a$; (3) a decoder D with parameters $\theta_D$, which receives the attacked video $V_a$ and reveals the secret message $M'$ from it. Although some methods use a discriminator to determine whether the video contains watermark information in order to improve invisibility, it is not an essential component. In the following, we provide a detailed explanation of each component and analyze why deep learning-based image watermarking methods can be applied to video watermarking.

\textbf{Encoder.}
The objective of the encoder is to embed the secret message into the cover video with minimal visual distortion. Therefore, the encoder should be approached with a thorough understanding of the pixel distribution of the cover video. As shown in Figure \ref{fig:framework}, cover video $V_c$ is fed into a cover video processor to extract the features of the cover video as the latent representations. There is a message processor to extend the one-dimensional message with the same shape as cover video latent representations in the spatiotemporal dimension. Afterward, the two latent representations are concatenated on the channel dimension and passed through the watermark generator. The purpose of this generator is to create a watermark that takes into account the pixel distribution of the cover video, thereby minimizing its impact on the video. The final step of the encoder is to combine the watermark and cover video through addition or a network to generate the watermarked video $V_w$. Hence, an ideal encoder is able to embed secret messages into the cover video based on the pixel distribution of the cover video. The loss function of encoder $\mathcal{L}_E$ is calculated by mean square error between $V_c$ and $V_w$,

\begin{equation}
    \mathcal{L}_E = MSE(V_c, V_w) = MSE(V_c,E(\theta_E, V_c, M)).
\end{equation}

Additionally, we propose a frame loss $\mathcal{L}_F$ to avoid significant differences in watermark strength added by the encoder on different frames in a video. Suppose that $F^{(i)}(\cdot)$ indicates the i-th frame of video, the frame loss $\mathcal{L}_F$ is defined as follows,
\begin{equation}
    \mathcal{L}_F = \sum_i^L \mathscr{l}_1^2(F^{(i)}(V_c), F^{(i)}(V_w)).
\end{equation}
Here, $\mathscr{l}_1$ refers to the $L_1$ norm function used for measuring the difference between two frames.


\textbf{Attack Simulation Layer.}
Adding distortions to watermarked video $V_w$ has been demonstrated as the most effective approach for enhancing watermark robustness \cite{zhu2018hidden,jia2021mbrs,zhang2021towards}. In order to optimize network parameters against various distortions, it is typically essential to incorporate a differentiable attack simulation layer (ASL) that is trained jointly with other fundamental modules. However, there are some distortions that are non-differential, like H.264 compression. The neural network cannot gain robustness enhancement from the non-differential distortion during training. In order to overcome such limitations, it is common practice to use a differentiable approximation to replace the non-differentiable distortion. Besides, Zhang et al. \cite{zhang2021towards} proposed forward ASL as the common solution for non-differential distortion. This method involves calculating the difference between attacked and watermarked video $V_a - V_w$, which is then referred to as pseudo-distortion. The pseudo-distortion is added to the watermarked video to generate a pseudo-attacked video. During backward propagation, the gradient from the decoder directly backpropagates to the encoder without passing through the ASL since the pseudo-distortion is set to not participate in gradient propagation. In this work, we utilize forward ASL to handle non-differentiable distortions. 

\textbf{Decoder.}
In the decoding process of watermarking, the decoder first extracts the watermark information by downsampling the attacked video $V_a$ and then reveals the message $M'$ according to the watermark information. The goal of the decoder is to minimize the prediction error of the message, so the loss function of the decoder $\mathcal{L}_D$ is defined as 

\begin{equation}
    \mathcal{L}_D = MSE(M, M`) = MSE(M, D(\theta_D, V_a)).
\end{equation}
In total, the target loss function $\mathcal{L}_{total}$ is defined as:
\begin{equation}
    \mathcal{L}_{total} = \lambda_E \mathcal{L}_E + \lambda_D \mathcal{L}_D + \lambda_F \mathcal{L}_F ,
\end{equation}
where $\lambda_E$, $\lambda_D$ and $\lambda_F$ are the weights for balancing the robustness and invisibility.

\section{Convolutional blocks for video}
Here, we discuss various video convolutional blocks in CNNs to explore the importance of temporal convolution and optimize the model size and computational cost for video watermarking applications.  

\subsection{2D convolutions}
2D CNNs for videos \cite{simonyan2014two} treat the L frames as channels of the video and merges the channel dimension and temporal dimension of the video. Therefore, the initial convolutional layer in the network compresses the temporal information of a video into channel feature maps, which prevents any subsequent temporal reasoning. This CNN architecture is depicted in Figure \ref{fig:conv}(a). 

\subsection{3D convlutions}
3D convolutions \cite{ji20123d}, also known as spatiotemporal convolutions, are a type of convolutional neural network (CNN) operation used for processing 3D data.
The convolutional kernels are convoluted in 3D, which means they are applied across both the temporal and spatial dimensions.
This type of CNN architecture is illustrated in Figure \ref{fig:conv} (b).

\subsection{(2+1)D convolutions}

(2+1)D convolutions \cite{tran2018closer} are a type of convolutional neural network layer that is commonly used for video processing. This layer combines 2D and 1D convolutions, hence the name (2+1)D. In this type of convolutional layer, a 2D convolutional kernel is applied spatially to each video frame separately. Then, a 1D temporal convolutional kernel is applied across the temporal dimension of the resulting feature maps. By using this approach, the network can capture spatial and temporal information separately within a video.
Compared to 3D convolutions, which directly operate on spatiotemporal data, (2+1)D convolutions have fewer parameters and are less computationally expensive. As such, they are often preferred for video analysis tasks on resource-limited platforms. This type of CNN architecture is illustrated in Figure \ref{fig:conv} (c).

\subsection{Depthwise convolutions}
Depthwise convolution \cite{howard2017mobilenets,tran2019video} is a specific convolutional neural network layer that applies individual kernel to each input channel in the input tensor. Depthwise convolutions are typically used in lightweight architectures, where the number of parameters and computation cost are critical.
In a depthwise convolutional layer, a pointwise convolution is applied to the input channels, which preserves all channel interactions.
Then, a convolutional kernel is applied depthwise (in the channel dimension) to each feature map separately. This means that each channel of the input feature map is convolved with a separate kernel. Depthwise convolutions can significantly reduce the number of parameters and the computation cost by decomposing the regular convolution into two smaller operations.
This makes them more memory efficient and faster to train. 
These types of convolutions are illustrated in Figure \ref{fig:conv} (d) and (e). Similar to 2D convolutional blocks, we merge the channel dimension and temporal dimension of the video in depthwise 2D convolutions.


\begin{table*}
  \caption{Comparison of PSNR and bit accuracy on a variety of common distortions for different watermarking methods. We utilize our \tool  to adapt MBRS and CIN to video watermarking. For Hidden and DVMark, we directly use results reported in \cite{luo2021dvmark}.}. 
  \label{tab:reslut}
  \resizebox{\linewidth}{!}{
  \begin{tabular}{c|ccccccccc}
     \hline
      \hline
    Methods &PSNR & \makecell[c]{H.264 \\ (CRF=22)} & \makecell[c]{Frame Average \\ (N=3)}& \makecell[c]{Frame Drop \\ (p=0.5)}& \makecell[c]{Frame Swap \\ (p=0.5)}& \makecell[c]{Gaussian Blur \\ ($\sigma$=2.0)}& \makecell[c]{Gaussian Noise \\ (std=0.04)} & \makecell[c]{Random Crop\\ (p=0.4)} & \makecell[c]{Random Hue \\ (p=1.0)}\\
          \hline
      &  \multicolumn{9}{c}{Kinetics-600} \\
      \hline
    Hidden & 37.00 & 79.85& 96.91 & 99.03 & 99.10 & 72.70 & 91.27  & 95.27 & 98.98 \\
    DVMark & 37.00 & 92.94& 98.10 & 98.99 & 99.35 & 98.09 & 98.56  & 97.06 & 99.81 \\
    \tool-$_{_{CIN}}$ & 37.13 & 98.44 & 99.99 & 99.76 & 99.94 & 99.28 & 99.99 & 98.81 & 99.89\\
    \tool-$_{_{MBRS}}$ & 37.97 & 99.74 & 99.99& 99.49& 99.54 & 99.99 & 99.99& 99.22& 99.99\\
     \hline
      &  \multicolumn{9}{c}{Inter4K} \\
     \hline
    \tool-$_{_{CIN}}$ & 37.03 & 97.66 & 99.99& 99.96& 99.45& 99.42 & 99.99& 99.27& 99.78 \\
    \tool-$_{_{MBRS}}$ & 37.97 & 99.67 & 99.99& 99.88& 99.51& 99.99 & 99.99& 99.74& 99.67 \\
     \hline
  \end{tabular}}
\end{table*}

\begin{table*}
  \caption{Comparison of performance and computational overhead of various convolutional blocks. We do not list the bit accuracy under frame average and Gaussian noise since they are all 99.99\%. The depthwise 2D convolutional block requires significantly fewer computational resources compared to other convolutional blocks yet delivers comparable performance.
  }
  \label{tab:conv}
  \begin{tabular}{c|c|cccccc|cc}
    \hline
    \hline
  \makecell[c]{Convolutional \\ blocks} &PSNR & \makecell[c]{H.264 \\ (CRF=22)} & \makecell[c]{Frame Drop \\ (p=0.5)}& \makecell[c]{Frame Swap \\ (p=0.5)}& \makecell[c]{Random Crop\\ (p=0.4)} & \makecell[c]{Gaussian Blur \\ ($sigma$=2.0)}& \makecell[c]{Random Hue \\ (p=1.0)} & \makecell[c]{Params \\ $\times 10^6$ }& \makecell[c]{FLOPs \\ $\times 10^9$ }\\
    \hline
    2D & 39.9  & 98.61& 99.58& 99.98& 99.60 & 99.99 & 99.82 & 20.8 & 15.57 \\
 3D & 37.1& 99.91 & 98.92 & 99.98&  97.27 & 99.76 & 99.86 & 54.2 & 234.63\\
    (2+1)D &  37.6& 99.61 & 99.97& 99.74 &  95.12  & 99.81 & 99.61 & 26.2 & 115.47 \\

         Depthwise 2D & 38.3& 99.22   & 99.61 & 99.68 & 99.99 & 99.99 & 99.92& 5.99 & 4.55\\
    Depthwise 3D & 38.5& 99.80   &99.98 & 99.93 & 99.99& 99.91& 99.76& 6.08 & 28.82 \\
    \hline
  \end{tabular}
\end{table*}

\section{Experiments}
\label{exp}

\subsection{Implementation Details}
Our model is implemented by Pytorch \cite{collobert2011torch7} and trained with NVIDIA A100 graphics cards, the batch size is set to 16, and the Adam optimizer with a learning rate $10^{-5}$ and default hyperparameters \cite{kingma2014adam}. In the training stage, each batch randomly selects a distortion from distortion set \{ Identity, H.264, Frame Average, Frame Drop, Frame Swap, Gaussian Blur, Gaussian Noise, Random Crop, Random Hue \}. In the evaluation stage, we use the trained model to evaluate the performance of each type of noise individually. During the training phase, we initially train the model without noise. At this point, we set the loss weights to $\lambda_E=1$, $\lambda_D=0.1$ and $\lambda_F=0$. Then, we proceed to train the model to resist different types of noise by loading the previously trained noise-free model and adjusting the loss weights to $\lambda_E=1$, $\lambda_D=0.01$ and $\lambda_F=0.05$, respectively.

\subsection{Dataset} 
To verify the robustness and invisibility of the adapted video watermarking method, we utilize the Kinetics-600 dataset \cite{carreira2017quo, carreira2018short} and Inter4K \cite{stergiou2022adapool} for training and evaluation.
 We train the models on 1000 randomly cropped video clips with dimension $8 \times 128 \times 128$ in the Kinetics-600 training set and evaluate the models on 1000 randomly selected video clips in the Kinetics-600 validation set. Because the resolution of the videos in the Kinetics-600 dataset is relatively low, we also use the Inter4K dataset to evaluate the generalization ability of models. The Inter4K dataset contains 1000 high-definition videos with a 4K resolution.
 For each input video clip, there is a corresponding secret message which is randomly sampled from the binary distribution $M~\{0,1\}^m$.

\subsection{Evaluation Metrics}
To objectively evaluate the robustness and invisibility of the models, we apply two quantitative metrics. To validate the robustness, we evaluate the accuracy between the original message $M$ and the revealed message $M'$. Accuracy is defined as follows:

\begin{equation}
    Accuracy(\%) = 1 - (\frac{1}{m} \times  |M - M'| \times 100\%)
\end{equation}

For the invisibility of watermarked videos, we utilize the Peak Signal-to-Noise Ratio (PSNR) \cite{almohammad2010stego} between cover video $V_c$ and watermarked video $V_w$ for evaluation.

\begin{equation}
    PSNR(V_c, V_w) = 20 \times log_{10} \frac{MAX(V_c,V_w)-1}{MSE(V_c,V_w)}
\end{equation}
where $MAX(\cdot)$ represents the maximum pixel value of videos, and $MSE(\cdot)$ is the mean square error.

\subsection{Comparison methods}
Our primary objective is to adapt deep learning-based image watermarking methods to video watermarking. So we use two state-of-the-art deep learning-based image watermarking methods \textbf{MBRS} \cite{jia2021mbrs} and \textbf{CIN} \cite{ma2022towards} to demonstrate the effectiveness of our \tool. Note that we use \tool-$_{_{MBRS}}$  and \tool-$_{_{CIN}}$ to represent the corresponding adapted video watermarking methods in this paper. The MBRS network is an auto-encoder that utilizes Squeeze-and-Excitation blocks \cite{hu2018squeeze}. On the other hand, CIN has used invertible neural networks for the first time in digital watermarking.
Then, we compare our models with two deep learning-based video watermarking methods \textbf{Hidden} \cite{zhu2018hidden} and \textbf{DVMark} \cite{luo2021dvmark}, especially with the state-of-the-art method DVMark proposed in \cite{luo2021dvmark}, which is robust against various distortions, such as H.264 compression, crop and Gaussian blur. For a fair comparison, we directly report the results listed in \cite{luo2021dvmark}.

Additionally, we are also interested in the impact of different convolutional blocks on video watermarking methods. 
So we compare the performance and computational cost of the \tool-$_{_{MBRS}}$  network with different convolutional blocks.

\begin{figure*}[ht]
  \centering
  \includegraphics[width=0.82\linewidth]{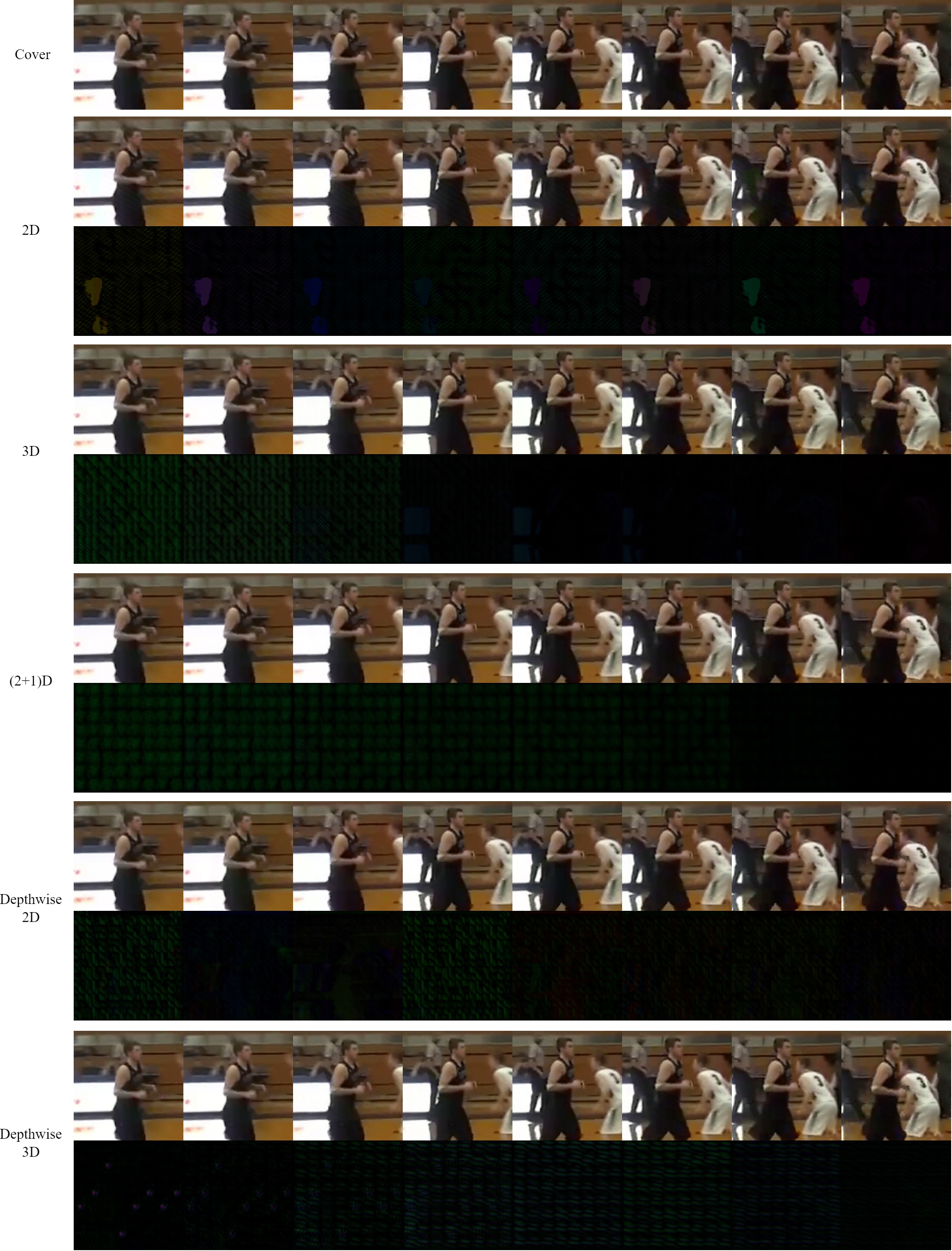}
  \caption{Visual samples of the cover and watermarked video sequence of frames from various convolutional blocks. From top to bottom: cover, 2D, 3D, (2+1)D, Depthwise 2D, Depthwise 3D. For each convolutional block, there are two rows of images. The top row is the watermark video frame, and the bottom row is the residual signal between the cover video and the corresponding watermarked video, magnified five times.}
    \label{fig:result}
\end{figure*}

\subsection{Evaluation}
In this section, we compare both robustness and invisibility of our method \tool against the baseline method Hidden and DVMark. The visual samples of watermarked video frames are shown in Figure \ref{fig:visual}.
It is evident that the cover video frames $V_c$ and watermarking video frames $V_w$ are visually indistinguishable, which indicates  that our two models adaptively embed the message into the cover video while maintaining high invisibility. Additionally, we can see the strength of the watermark varies depending on the complexity of the region from the residual signal.
To evaluate the robustness, we measure the bit accuracy of each model on a large collection of common distortions: H.264 compression (CRF = 22), frame average (N = 3), frame drop (p = 0.5), frame swap (p = 0.5), Gaussian blur ($\sigma$ = 2.0), Gaussian noise (std = 0.04), random crop (p = 0.4), random hue (p = 1.0).
All results are evaluated on video clips of size $8\times 128 \times 128$, with the message length m = 96. 

As shown in Table \ref{tab:reslut}, the adapted methods \tool-$_{_{CIN}}$  and \tool-$_{_{MBRS}}$  have higher PSNR values, and their bit accuracy under various distortions surpasses that of Hidden and DVMark. Especially under H.264 compression, the bit accuracy of \tool-$_{_{CIN}}$  and \tool-$_{_{MBRS}}$  is much higher than that of DVMark, with differences of 6.8\% and 5.5\%, respectively, despite using different H.264 compression training methods. 
Additionally, it's significant to mention that \tool-$_{_{MBRS}}$  has a bit accuracy of over 99\% in all of the distortions in Table \ref{tab:reslut} simultaneously. This makes it a strong contender for practical implementation.

\textbf{Generalization ability.} The \tool-$_{_{CIN}}$ and \tool-$_{_{MBRS}}$  models, trained solely on the Kinetics-600 dataset, both produce excellent results on Inter4K datasets as shown in Table \ref{tab:reslut}. This highlights the exceptional generalization ability of these two models, which is crucial for practical applications.

\subsection{Comparison of various convolutional blocks}
In order to explore the impact of different convolutional blocks on video watermarking, we modify the convolutional blocks and test their performance and computational complexity for video watermarking tasks. Here, we use \tool-$_{_{MBRS}}$ as an example. For a fair comparison, all the models are trained in 2000 epochs. 
The results of \tool-$_{_{MBRS}}$ with various convolutional blocks are visualized in Figure \ref{fig:result}. For each type of convolutional block, we show the watermarked video frame sequence and corresponding residual signal. We can see the networks with different convolutional blocks have different strategies for embedding watermarks. The watermark pattern of (2+1)D seems to be independent of the cover video. And the network with 3D convolutional blocks embeds a strong watermark in the first two frames of the video. The remaining three types of convolutional blocks can evenly embed watermark information into all frames of the cover video.

Table \ref{tab:conv} presents results on 1000 video clips in the Kinetics-600 validation set. 
There are some insights inferred from these results. First, there is a noticeable gap between the bit accuracy under the random crop of the 2D block and (2+1)D block. This indicates that the introduction of temporal convolution in video watermarking networks will lead to a decrease in performance, which is likely due to the complex convolutional block being more prone to overfitting. Besides, we find that using depthwise convolutions significantly decreases the number of parameters and FLOPs in our model, especially when utilizing 3D convolution. However, the overall performance of the model is minimally affected as pointwise convolutions preserve channel interactions.
This phenomenon also indicates the problem of over-parameterization in regular 2D/3D convolutions in digital watermarking networks.

\begin{figure}[h]
  \centering
  \includegraphics[width=\linewidth]{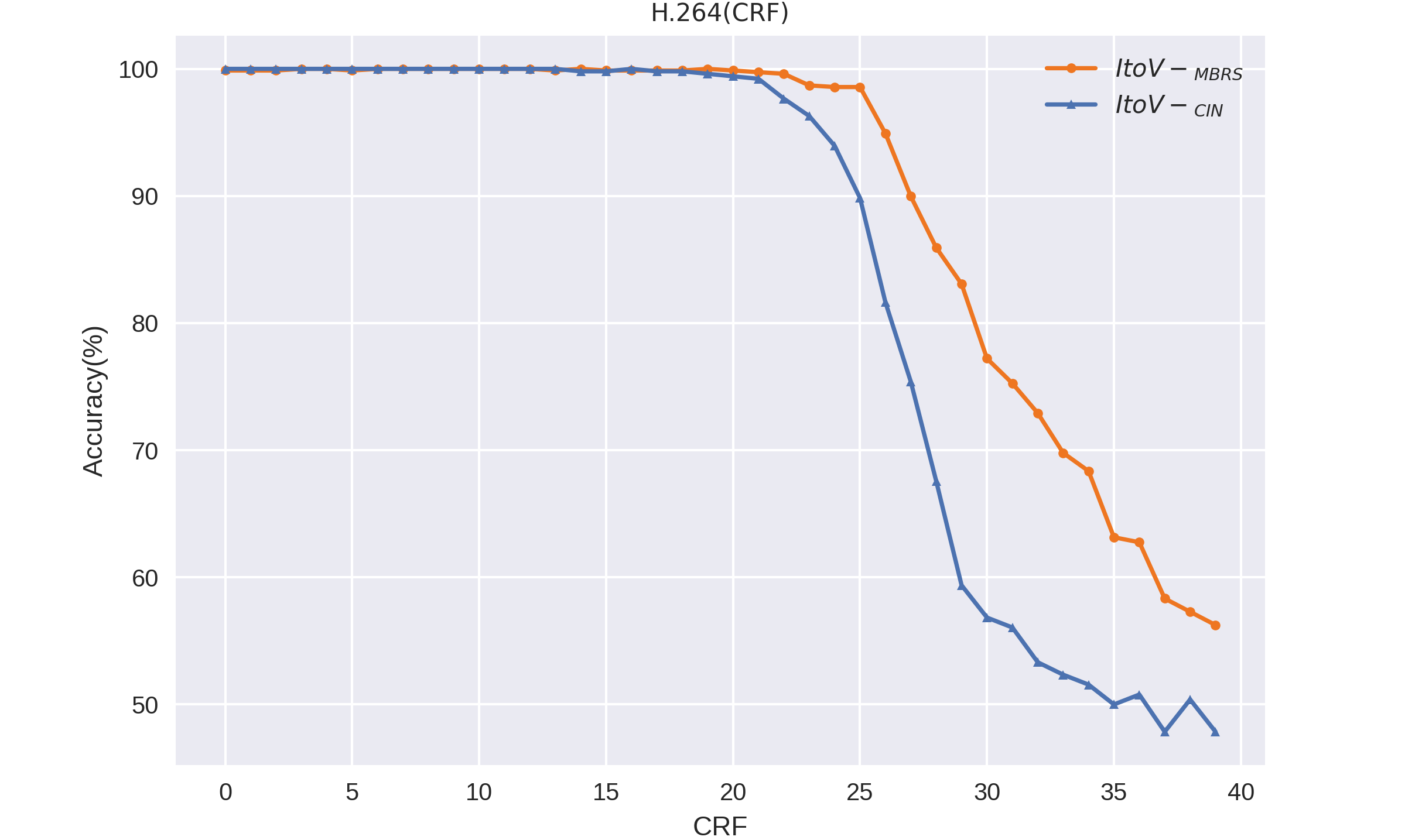}
  \caption{Bit accuracy of \tool-$_{_{MBRS}}$  and \tool-$_{_{CIN}}$ for H.264 compression with various constant rate factors (CRF) on Kinetics-600 validation set.}
    \label{fig:crf}
\end{figure}

\subsection{H.264 compression and screen recording}

Screen recording is a prevalent form of video infringement in the real world. With the development of Internet streaming media, software-based screen recording is more common than camera-based screen recording due to its simplicity and effectiveness.
 In Internet video sharing platforms, video compression parameters are usually expressed in bitrate rather than CRF to meet the streaming bandwidth limitations. The bitrate is fixed, so the quality may vary depending on the video content. Complex scenes will be compressed more, resulting in lower quality. Therefore, H.264 compression algorithms with fixed bitrate will adjust the compression rate (CRF value) according to the video scene. In order to explore the stability of watermarking methods against H.264 compression, we test the bit accuracy of \tool-$_{_{CIN}}$  and \tool-$_{_{MBRS}}$  models under H.264 compression with different CRF values. Note that here we use the models mentioned in the previous section that can resist various distortions simultaneously rather than just resisting H.264 compression. As shown in Figure \ref{fig:crf}, \tool-$_{_{MBRS}}$  has better robustness than \tool-$_{_{CIN}}$  for H.264. When the CRF value is less than 20, the bit error rate of the \tool-$_{_{MBRS}}$  model is almost 0. Even in the CRF value range of 20 to 25, the \tool-$_{_{MBRS}}$  model can still maintain excellent bit accuracy. After the CRF value exceeds 25, the accuracy of the \tool-$_{_{MBRS}}$  model begins to decrease rapidly.
In addition, we also test the bit accuracy of \tool-$_{_{MBRS}}$  and \tool-$_{_{CIN}}$  in software screen recording under real-world scenarios. We first live-stream our watermarked videos at a bitrate of 1000kbps and then use OBS studio\cite{obs} with the default setting for screen recording. In this case, the average bit accuracy for \tool-$_{_{MBRS}}$  and \tool-$_{_{CIN}}$  are 97.86\% and 96.92\%, respectively. The results are consistent with our expectations because the screen recording can be regarded as the decoding and re-encoding process of the H.264 algorithm \cite{jin2022streamingtag}. As previously mentioned, our networks provide an effective defense against such compound distortions.

\subsection{Ablation Study}

\textbf{Effectiveness of $\mathcal{L}_F$ loss.} The $\mathcal{L}_F$ loss is designed to guarantee that the watermark intensity in each frame of the video clip is consistent so that there are no instances of low visual quality in any frame. We measure the PSNR values of each frame in the watermarked video generated by \tool-$_{_{MBRS}}$ with and without $\mathcal{L}_F$ loss, and the results are listed in Table \ref{tab:frame}. We can see that the $\mathcal{L}_F$ loss significantly reduces the difference in watermark intensity between different frames. Specifically, the standard deviation reduces significantly from 1.344 dB to 0.159 dB. Although the average PSNR value slightly decreases, it is important to note that the human eye is more sensitive to individual low-quality frames. With the implementation of $\mathcal{L}_F$ loss, there is an improvement in the lowest PSNR value from 38.43 dB to 39.63 dB, indicating that this loss can effectively enhance the invisibility of video watermarking.

\begin{table}[h]
  \caption{The PSNR value of \tool-$_{_{MBRS}}$ with and without $\mathcal{L}_F$ loss. The two models achieve similar robustness.
  Here, f1 represents the first frame of the video clip, and so on for the following. }
  \label{tab:frame}
  \resizebox{\linewidth}{!}{
  \begin{tabular}{c|cccccccc|cc}
\hline
      \hline
$\mathcal{L}_F$ loss & f1 & f2 & f3 & f4 & f5 & f6 & f7 & f8 & mean & std \\

\hline
 \XSolidBrush  & 41.10 & 41.92 & 38.45 & 40.78 & 38.43 & 40.47 & 38.44 & 41.23 & 40.10 & 1.344 \\
 \Checkmark   & 40.07 & 40.05 & 39.99 & 40.02 & 39.88 & 39.79 & 39.69 & 39.63 & 39.89 & 0.159 \\
     \hline
  \end{tabular}}
\end{table}



\section{Conclusion}
In this paper, we have presented an insight into deep learning-based video watermarking, which is that watermarking neural network can treat images and videos equally because they are just different shapes of data for watermarking.
Inspired by this insight, we have proposed a method \tool to adapt deep learning-based image watermarking to video watermarking efficiently. As a result, the state-of-the-art deep learning-based image watermarking method can be efficiently used for video watermarking. Furthermore, we have investigated the impact of spatial convolution, temporal convolution, and depthwise convolution on video watermarking. Our findings indicate that spatial convolution has a more significant influence than temporal convolution in this process. Besides, we have found that incorporating depthwise convolutions can significantly decrease computational costs while maintaining performance. In network training, we have designed a new frame loss to embed watermark information more evenly into each video frame. Extensive experiments have shown that our approach \tool could achieve better performance in robustness and invisibility than previous video watermarking methods.

\bibliographystyle{ACM-Reference-Format}
\bibliography{sample-authordraft}










\end{document}